\documentclass[]{article}
\usepackage[a4paper, margin=1in]{geometry}
\usepackage{authblk}	
\usepackage{amsmath}    
\usepackage{graphicx}	
\usepackage{caption}	
\usepackage{subcaption}	
\usepackage[flushleft]{threeparttable}	
\usepackage{multirow}	
\usepackage{amsfonts}   
\usepackage{amssymb}	
\usepackage{amsmath}	

\usepackage[
	backend=bibtex,
	style=phys,
	sorting=none
]{biblatex}
\title{Multifractal behaviour in multiparticle production in $pp$ collisions at $\sqrt{s}=$ 0.9, 7 and 8 TeV from the CMS experiment}

\author{Z. Ong, P. Agarwal, H.W. Ang, A.H. Chan, C.H. Oh}
\affil{Department of Physics, National University of Singapore}
\date{}

\begin{document}

\maketitle

\begin{abstract}
	Multifractal analysis was performed on $pp$ collision data at $\sqrt{s}=$ 0.9, 7 and 8 TeV from the CMS experiment at CERN. The data was obtained and processed from the CERN Open Data Portal. Vertical analysis was used to compute the generalised dimensions $D_q$ and the multifractal spectra $f(\alpha)$ of the data, which reveals the level of complexity of its pseudorapidity distribution. It was found that the $f(\alpha)$ curves widen with increasing collision energy, signalling an increase in branching complexity.
\end{abstract}

\section{Multifractal formalism for multiparticle production}

Multifractal analysis is a powerful tool used to characterise the complexity of data. It is highly multi-disciplinary in nature, finding applications in the analysis of a wide variety of complex systems, such as studying tectonic processes~\cite{TELESCA2003385}, medical signal analysis~\cite{lopes_fractal_2009} and time series analysis in meteorology~\cite{Krzyszczak2019}. Here, we re-introduce a formalism tailored for investigating multiparticle production, as formulated by Hwa~\cite{Hwa:1989vn}. As with the original formulations, rapidity $y$ will be used in the presentation, but will be substituted with pseudorapidity $\eta$ when processing the data.

\subsection*{\textit{G}-moments}

Multifractal analysis in the context of multiparticle production was originally done by computing the \textit{G}-moments of multiplicity distributions~\cite{Hwa:1989vn,Chen:1993fg,DeWolf:1995nyp}. We begin by considering a single rapidity interval $\delta y$ and a collection of $N$ events. Each event $i$ has multiplicity $n_i$, and $K = \sum_{i=1}^{N} n_i$ is the total number of particles in $\delta y$ summed over all events.

The \textit{G}-moment for the collection of $N$ events in $\delta y$ is then defined as
\begin{equation}
	G_q (\delta y) = {\sum_{i=1}^{N}} {'p_i}^q,
	\label{eq:G-moments}
\end{equation}
where $p_i = \frac{n_i}{K}$ is the probability of finding a particle in the $i$\textsuperscript{th} event. The prime indicates $G_q$ is summed only over non-empty events (i.e. all $p_i > 0$), so we can have $q \in \mathbb{R}$. However, $G_q (\delta y)$ is very sensitive to statistical fluctuations for small $N$, and modified definitions have been proposed in~\cite{Hwa:1991rd}.

The quantity being summed over ($p_i$ in our case) is called a \textit{measure}. $q$ acts as a probing parameter -- higher values would enhance the differences between the $p_i$'s. Changing $\delta y$ explores the phase space at different scales. As such, $G_q (\delta y)$ is sometimes called the partition function in multifractal analysis~\cite{TELESCA2003385} as it encodes information at different scales $\delta y$ and at different moments $q$. This scheme of setting up $G_q (\delta y)$ is analogous to the box counting algorithm commonly used in digital image analysis.

\subsection*{Properties of $G_q (\delta y)$}

Suppose that for some event $i$, its $p_i$ value scales as
\begin{equation}
	p_i \propto (\delta y)^{\alpha_i}
	\label{eq:BinScaling}
\end{equation}
for some exponent $\alpha_i$. For a multifractal system, the $\alpha_i$'s are generally different and can take on a range of values, reflecting different subsets of the data with local scaling behaviour\footnote{If all $\alpha_i$'s are equal, the system is a monofractal.}. Let $S_{\alpha'}$ be a set containing all the events that scale according to equation~\ref{eq:BinScaling}, with the same value of $\alpha'$. Furthermore, let $N_{\alpha'} (\delta y)$ denote the cardinality of $S_{\alpha'}$. Multifractals generally have the property that~\cite{Hwa:1989vn,TELESCA2003385}
\begin{equation}
	N_{\alpha'} (\delta y) \propto (\delta y)^{-f(\alpha')},
\end{equation}
i.e. the collection of events that scale according to equation~\ref{eq:BinScaling} with a particular value of $\alpha'$ in a particular window $\left[ \alpha', \alpha' + d\alpha' \right]$ form a fractal subset (of a larger set of more fractal subsets corresponding to other values of $\alpha$), and its cardinality scales with fractal dimension $f(\alpha')$. It is in this sense that the union of fractal subsets form a multifractal.

$N_{\alpha'} (\delta y)$ has been suggested by Halsey \textit{et al.}~\cite{Halsey:1986zz} to be of the form~\cite{Hwa:1989vn}
\begin{equation}
	N_{\alpha'} (\delta y) = \int_{\alpha'}^{\alpha' + d\alpha'} d \alpha h(\alpha) (\delta y)^{-f(\alpha)},
	\label{eq:Halsey}
\end{equation}
for some continuous function $h(\alpha)$. With equation~\ref{eq:Halsey}, we can now re-express $G_q (\delta y)$ as
\begin{equation}
	\begin{split}
		G_q (\delta y) &= {\sum_{m=1}^{M}} {'p_m}^q\\
		&\propto {\sum_{m=1}^{M}} {'(\delta y)}^{q\alpha_m}\\
		&= \int d \alpha h(\alpha) (\delta y)^{-f(\alpha)} (\delta y)^{q\alpha}\\
		&= \int d \alpha h(\alpha) (\delta y)^{q\alpha - f(\alpha)}.
	\end{split}
	\label{eq:Gq_integral}
\end{equation}
The resulting dependence of $G_q (\delta y)$ on $\delta y$ can be expressed as~\cite{Hwa:1989vn}
\begin{equation}
	G_q (\delta y) \propto (\delta y)^{\tau (q)},
	\label{eq:G-moment-scaling-law}
\end{equation}
where $\tau (q)$ has been established by Hentschel and Procaccia~\cite{Hentschel:1983zhc} to be
\begin{equation}
	\tau (q) = (q-1) D_q.
	\label{eq:HentschelProcaccia}
\end{equation}
$D_q$ is known as the generalised dimension of order $q$, which can be experimentally obtained by rearranging equation~\ref{eq:HentschelProcaccia}:
\begin{equation}
	D_q = \lim\limits_{\delta y \rightarrow 0} \left[ \frac{1}{q-1} \frac{\ln G_q (\delta y)}{\ln (\delta y)} \right].
	\label{eq:D_q}
\end{equation}
It must be noted that in experimental measurements, the mathematical limit $\delta y \rightarrow 0$ in equation~\ref{eq:D_q} cannot be realised. The finiteness of particle multiplicity produced from finite energy implies that self-similar and fractal structures, if present, cannot persist indefinitely to finer scales of resolution~\cite{Florkowski:1990ba}. Additionally, the detector resolution also imposes a lower limit on the probing scale.


The goal in multifractal analysis is to study the dependence of $G_q (\delta y)$ on $\delta y$, and the $D_q$'s are the main quantities that summarise the relation. $D_0$, $D_1$ and $D_2$ are also known as the \textit{fractal}, \textit{information}\footnote{For $q=1$, the singularity in $1/(q-1)$ is handled by taking the limiting value as $q \rightarrow 1$.} and \textit{correlation} dimensions respectively~\cite{Hentschel:1983zhc,Hwa:1989vn}. A monofractal has constant $D_q$ for all $q$; otherwise, it is a multifractal~\cite{TELESCA2003385}.

The above approach using the \textit{G}-moments does not assume any specific dynamical model of multiparticle production~\cite{Florkowski:1990ba}, which serves as a model-agnostic tool to describe the complexity within data.

\section{Vertical and horizontal averaging}

By construction, the \textit{G}-moment described in the previous section accesses a very small subset of the data -- only one rapidity interval over all $N$ events. The \textit{G}-moments computed this way is also known as the vertical moments~\cite{Hwa:1989vn}, notated as $G_q^\text{(v)}$.

Alternatively, one can also analyse all the rapidity intervals in a single event. The \textit{G}-moments computed this way is known as the horizontal moments~\cite{Hwa:1989vn}, notated as $G_q^\text{(h)}$.

Since both methods access only a small slice of the available data, the statistics can be enhanced by supplementing the vertical moments with horizontal averaging, and vice versa. For example, the vertical moments can be calculated for every $\delta y$ interval and averaged horizontally over the $M$ bins:
\begin{equation}
	\left< G_q^\text{(v)} \right> \equiv \frac{1}{M} \sum_{m}^{M} G_q^\text{(v)}.
	\label{eq:VerticalG-momentsWithHorizontalAveraging}
\end{equation}

Likewise, the horizontal moments can be calculated for every event and averaged vertically over all $N$ events:
\begin{equation}
	\left< G_q^\text{(h)} \right> \equiv \frac{1}{N} \sum_{i}^{N} G_q^\text{(h)}.
	\label{eq:HorizontalG-momentsWithVerticalAveraging}
\end{equation}

In general, we have $\left< G_q^\text{(v)} \right> \neq \left< G_q^\text{(h)} \right>$ except when $q=1$. The two moments also capture different features of the dataset; for example, $\left< G_q^\text{(v)} \right>$ would be sensitive to rare events with very high multiplicity, while $\left< G_q^\text{(h)} \right>$ would not. $\left< G_q^\text{(h)} \right>$ however, describes a more intuitive notion of fractal structures within the multiparticle production in each event. Florkowski and Hwa~\cite{Florkowski:1990ba} have studied the limiting scenarios in which they are equivalent, under the assumptions of ergodicity.

Contemporary multiplicity measurements (e.g.~\cite{Khachatryan:2010nk}) are statistically derived quantities, with $P(n)$ interpreted as an average over many events that has undergone an unfolding process. The pseudorapidity distribution of single events is not available as a result. Hence, we will perform our multifractal analysis using the vertical moments with horizontal averaging (equation~\ref{eq:VerticalG-momentsWithHorizontalAveraging}).

\section{The $f(\alpha)$ spectrum}

As multifractals cannot be described by a single fractal dimension, a \textit{singularity spectrum} or \textit{Legendre spectrum}, $f(\alpha)$ is used to characterise them instead, which encodes the spread of $\alpha$ values exhibited by the system.

To obtain the $f(\alpha)$ spectrum, consider again equation~\ref{eq:Gq_integral}:
\begin{equation*}
	G_q (\delta y) = \int d \alpha h(\alpha) (\delta y)^{q\alpha - f(\alpha)}.
\end{equation*}
Suppose $h(\alpha) \neq 0$. For each value of $q$ and in the limit $\delta y \rightarrow 0$, the integral would have most of its contribution from some value $\alpha$ which makes the exponent $q\alpha - f(\alpha)$ smallest. Let this optimising value of $\alpha$ be $\alpha_q$. To minimise the exponent, we require
\begin{equation}
	\left. \frac{d}{d\alpha} \left[q\alpha - f(\alpha) \right] \right|_{\alpha=\alpha_q} = 0,
\end{equation}
\begin{equation}
	\left. \frac{d^2}{d\alpha^2} \left[q\alpha - f(\alpha) \right] \right|_{\alpha=\alpha_q} > 0,
\end{equation}
which result in
\begin{equation}
	f'(\alpha_q) = q,
	\label{eq:f-prime}
\end{equation}
\begin{equation}
	f''(\alpha_q) < 0.
	\label{eq:f-doublePrime}
\end{equation}
Substituting equations~\ref{eq:f-prime} and~\ref{eq:f-doublePrime} into~\ref{eq:D_q}, and using the saddle point approximation, we get
\begin{equation}
	\begin{split}
		D_q &= \lim\limits_{\delta y \rightarrow 0} \left[ \frac{1}{q-1} \frac{\ln G_q (\delta y)}{\ln (\delta y)} \right]\\
		&= \lim\limits_{\delta y \rightarrow 0} \left[ \frac{1}{q-1} \frac{\ln \int d \alpha h(\alpha) (\delta y)^{q\alpha - f(\alpha)}}{\ln (\delta y)} \right]\\
		&\approx \lim\limits_{\delta y \rightarrow 0} \left[ \frac{1}{q-1} \frac{\ln \left[d \alpha h(\alpha_q) (\delta y)^{q\alpha_q - f(\alpha_q)}\right]}{\ln (\delta y)} \right]\\
		&= \lim\limits_{\delta y \rightarrow 0} \left[ \frac{1}{q-1} \frac{\ln \left[d \alpha h(\alpha_q) \right] + \ln (\delta y)^{q\alpha_q - f(\alpha_q)}}{\ln (\delta y)} \right]\\
		\therefore D_q &\approx \frac{q\alpha_q - f(\alpha_q)}{q-1}
		= \frac{\tau(q)}{q-1}.
	\end{split}
	\label{eq:D_q_applied}
\end{equation}

Equation~\ref{eq:D_q_applied} reveals how all the quantities in multifractal analysis relate to each other and how they can be computed. First, $\tau(q)$ can be evaluated by~\cite{Hwa:1989vn}
\begin{equation}
	\tau(q) = \lim\limits_{\delta y \rightarrow 0} \left[ \frac{\ln G_q (\delta y)}{\ln (\delta y)} \right],
\end{equation}
which also gives $D_q$ via equation~\ref{eq:D_q}. In our analysis, $G_q(\delta y)$ will be replaced by $\left< G_q^\text{(v)} \right>$ (equation~\ref{eq:VerticalG-momentsWithHorizontalAveraging}).

Next, $\alpha_q$ is obtained via numerical differentiation~\cite{Hwa:1989vn}:
\begin{equation}
	\alpha_q = \frac{d}{dq} \tau(q).
\end{equation}
This finally allows us to compute $f(\alpha)$~\cite{Hwa:1989vn}:
\begin{equation}
	f(\alpha_q) = q \alpha_q - \tau(q).
	\label{eq:fa-applied}
\end{equation}

For a multifractal, equations~\ref{eq:f-prime} and~\ref{eq:f-doublePrime} indicate that the curve $f(\alpha)$ has a maximum at $\alpha_0$ and is concave downward everywhere. In the case of a monofractal, only a single value of $\alpha$ exists and the $f(\alpha)$ spectrum would reduce to a point~\cite{TELESCA2003385}.

\subsection*{What does the $f(\alpha)$ curve tell us?}

In general, $f(\alpha)$ curves in multifractal analysis describes the \textit{complexity} of a signal~\cite{TELESCA2003385}. In the context of our analysis, it describes the smoothness (or roughness) of the pseudorapidity distribution of our multiplicity data, $N(\eta)$. A low value of $\alpha_0$ indicates a smoother $N(\eta)$. The width $W$ reflects the range of fractal exponents embedded in the dataset -- larger values of $W$ reflect a more complex $N(\eta)$ distribution (i.e. greater degree of multifractality).

\section{About the data}

This analysis is performed on Run 1 data from the CMS collaboration processed from the CMS Open Data Portal, covering centre-of-mass energies $\sqrt{s}=$ 0.9, 7 and 8 TeV. The analysis method follows largely that of CMS~\cite{Khachatryan:2010nk}, which analysed minimum-bias (MinBias), non-single diffractive (NSD) multiplicity distributions.

NSD events were selected by requiring that at least one forward hadron (HF) calorimeter tower on
each side of the detector have at least 3 GeV of energy deposited in the event. The primary vertex was chosen as the vertex with the highest number of associated tracks, which must also be within 15 cm of the reconstructed beamspot in the beam axis and be of good reconstruction quality (ndof $>$ 4).

Good quality tracks were selected by requiring them to carry the \texttt{highPurity} label. Furthermore, we select for tracks with $<$10\% relative error on the transverse momentum ($p_{\text{T}}$) measurement ($\sigma_{p_{\text{T}}} / p_{\text{T}} < 0.1$) to reject low-quality and badly reconstructed tracks. Secondaries were removed by requiring a small impact parameter with respect to the selected primary vertex. Also, tracks were required to have $p_{\text{T}} > 500$ MeV/c, which will be extrapolated to zero via unfolding.

Finally, unfolding was performed using an iterative ``Bayesian unfolding method'', which is more accurately known as ``D’Agostini iteration with early stopping'' and described in~\cite{DAgostini:1994fjx}. This infers the original charged hadron multiplicity distribution (MinBias NSD) from the charged track multiplicity distribution measured.

Tables~\ref{tab:RECOdatasets} and~\ref{tab:MCdatasets} in the Appendix summarise the datasets used.

\section{Results and Discussion}

Table~\ref{tab:MultifractalAnalysisResults} summarises the results of our multifractal analysis, giving the generalised dimensions $D_q$ and an approximate description of the width of the $f(\alpha)$ curves.

\begin{table}[h]
	\centering
	\caption{Generalised dimensions and width of $f(\alpha)$ curves from multifractal analysis of $pp$ collisions at $\sqrt{s}=$ 0.9, 7 and 8 TeV, with all values normalised to $D_0$ at $\sqrt{s}=$ 900 GeV. $\alpha_\text{max}$ and $\alpha_\text{min}$ are evaluated at $q=-20$ and $q=+20$ respectively.}
	\begin{tabular}{ cccc }
		\hline
		$\sqrt{s}$ & 900 GeV & 7 TeV & 8 TeV \\
		\hline
		$D_0$	& 1		& 0.996 & 0.996 \\
		$D_1$	& 0.988 & 0.980 & 0.978 \\
		$D_2$	& 0.975 & 0.965 & 0.961 \\
		$D_3$	& 0.964 & 0.952 & 0.948 \\
		$D_4$	& 0.954 & 0.941 & 0.937 \\
		$D_5$	& 0.946 & 0.932 & 0.927 \\
		\hline
		$\alpha_\text{min}$	& 0.838 & 0.829 & 0.826 \\
		$\alpha_\text{max}$	& 1.054 & 1.068 & 1.072 \\
		$W$					& 0.216 & 0.239 & 0.246 \\
		\hline
	\end{tabular}	
	\label{tab:MultifractalAnalysisResults}
\end{table}

\begin{figure}[h]
	\centering
	\includegraphics[width=0.8\textwidth]{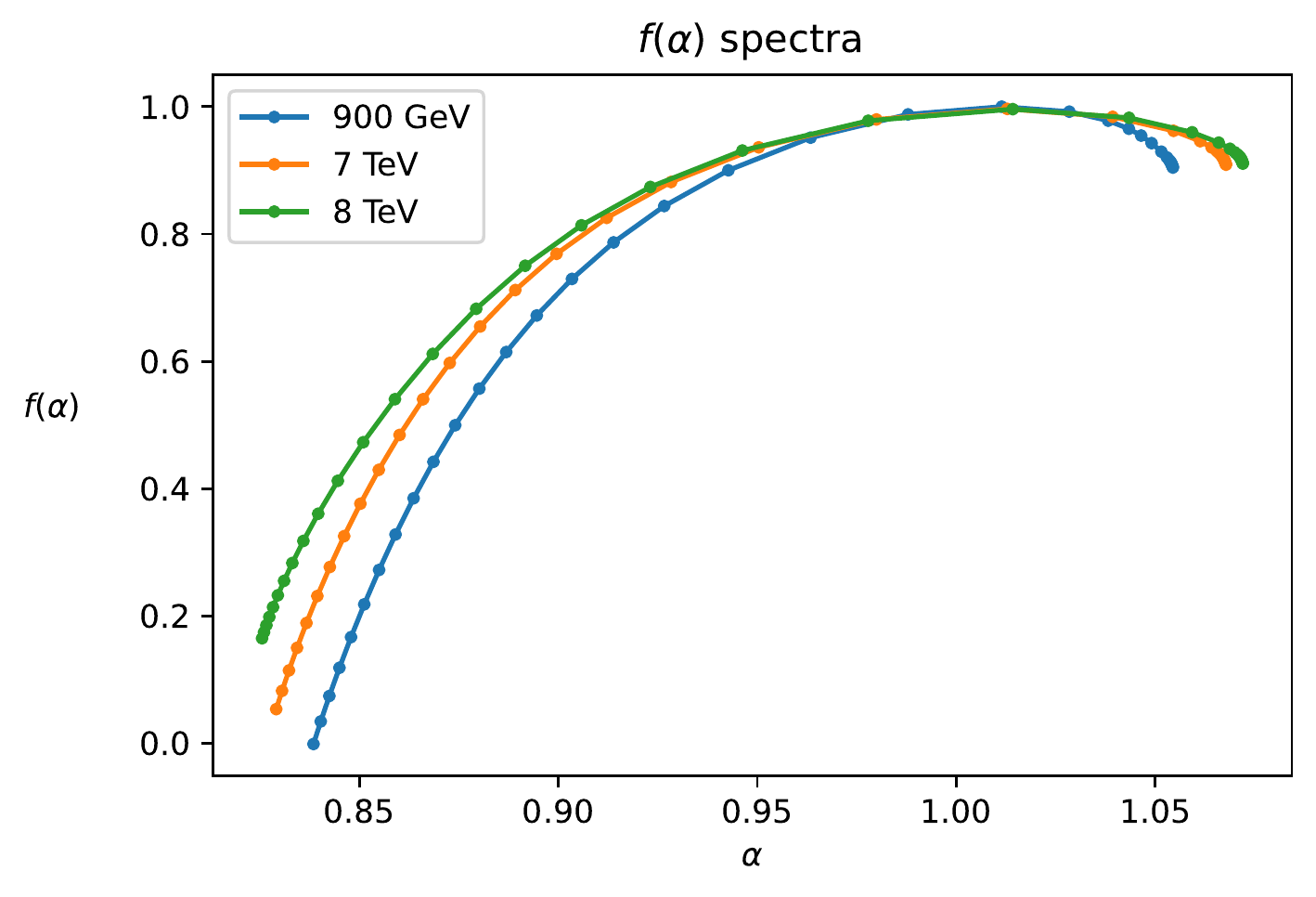}
	\caption{$f(\alpha)$ spectra of $pp$ collisions at $\sqrt{s}=$ 0.9, 7 and 8 TeV.}
	\label{fig:f(a)curves}
\end{figure}

Mathematically, the data points that constitute the $f(\alpha)$ curves are plotted by evaluating equation~\ref{eq:fa-applied} for $q \in \mathbb{R}$. However, this is computationally impossible, as equation~\ref{eq:G-moments} would produce infinities for $q \rightarrow -\infty$ and infinitesimally small values for $q \rightarrow +\infty$; both scenarios would lead to numerical instabilities. Since the goal is to obtain a relative comparsion of the widths of the $f(\alpha)$ curves, we restricted the computation of data points to $-20 \leq q \leq 20$. The distance between these data points at these limiting values of $q$ in the $\alpha$-axis would be our estimate\footnote{Some studies (e.g.~\cite{TELESCA2003385}) estimate $W$ by performing a simple fit of $f(\alpha)$ to a quadratic function and taking the distance between the roots. However, this assumes that $f(\alpha)$ is inherently quadratic, which is not always true.} of the $f(\alpha)$ width, $W$. 

Figure~\ref{fig:f(a)curves} shows that the $f(\alpha)$ curves clearly broaden with increasing collision energy (which is also detailed in Table~\ref{tab:MultifractalAnalysisResults}). This is a reflection of increasing complexity in the multiplicity data.

\section{Conclusion}

The techniques in multifractal analysis provide a model-independent means of describing the inherent complexity within the structures of the multiplicity distribution. We have used it to analyse $pp$ collisions at $\sqrt{s}=$ 0.9, 7 and 8 TeV and found that the $f(\alpha)$ curves broaden with increasing collision energy. This reflects an increase in complexity of the pseudorapidity distribution of the data. At higher energies, we would expect the $f(\alpha)$ curves to broaden further.

\newpage
\appendix

\section{Datasets used}

\begin{table}[h!]
	\centering
	\caption{Summary of CMS collider datasets used from CERN Open Data Portal}
	\begin{tabular}{ |c|l|c| } 
		\hline
		$\sqrt{s}$  & \multirow{2}{*}{Dataset} & \multirow{2}{*}{Ref.} \\
		(TeV) & & \\
		\hline \hline
		0.9	& /MinimumBias/Commissioning10-07JunReReco\_900GeV/RECO & \cite{data900GeV} \\
		\hline
		7	& /MinimumBias/Run2010A-Apr21ReReco-v1/AOD & \cite{data7TeV} \\ 
		\hline
		8	& /MinimumBias/Run2012B-22Jan2013-v1/AOD & \cite{data8TeV} \\ 
		\hline	
	\end{tabular}	
	\label{tab:RECOdatasets}
\end{table}

\begin{table}[h!]
	\centering
	\caption{Summary of Monte Carlo datasets used from CERN Open Data Portal}
	\begin{tabular}{ |c|l|c| } 
		\hline
		$\sqrt{s}$  & \multirow{2}{*}{Dataset} & \multirow{2}{*}{Ref.} \\
		(TeV) & & \\
		\hline \hline
		
		\multirow{2}{*}{0.9}	& /MinBias\_TuneZ2\_900GeV\_pythia6\_cff\_py & \multirow{2}{*}{\cite{mc900GeV}}\\
		& \_GEN\_SIM\_START311\_V2\_Dec11\_v2 &  \\
		\hline
		
		\multirow{2}{*}{7}	& /MinBias\_TuneZ2star\_7TeV\_pythia6/Summer12-LowPU2010 & \multirow{2}{*}{\cite{mc7TeV}} \\ 
		& \_DR42-PU\_S0\_START42\_V17B-v1/AODSIM &  \\ 
		\hline
		
		\multirow{2}{*}{8}	& /MinBias\_TuneZ2star\_8TeV-pythia6/Summer12\_DR53X-PU & \multirow{2}{*}{\cite{mc8TeV}} \\
		& \_S10\_START53\_V7A-v1/AODSIM & \\
		\hline	
	\end{tabular}	
	\label{tab:MCdatasets}
\end{table}

\printbibliography

\end{document}